\documentclass[twocolumn,superscriptaddress,longbibliography,aps,prl,preprintnumbers]{revtex4-2}

\usepackage[normalem]{ulem}
\usepackage{graphicx}
\usepackage{bm}
\usepackage{color}
\usepackage{epstopdf}
\usepackage{amsmath}
\usepackage{amssymb}
\usepackage{epstopdf}
\usepackage{lipsum}

\usepackage{gensymb}

\usepackage{multirow}

\usepackage{scrextend}

\usepackage[urlcolor=blue,colorlinks=true,citecolor=blue,linkcolor=blue,pdfstartview={FitH},bookmarks=false]{hyperref}

\usepackage[dvipsnames]{xcolor}

\graphicspath{{fig/}{./fig/}{.}}

\begin{document}

\title{
Comment on ``Fully gapped superconductivity and topological aspects of the noncentrosymmetric superconductor TaReSi''
}

\author{Andrzej~Ptok}
\email[e-mail: ]{aptok@mmj.pl}
\affiliation{\mbox{Institute of Nuclear Physics, Polish Academy of Sciences, W. E. Radzikowskiego 152, PL-31342 Krak\'{o}w, Poland}}

\date{\today}

\begin{abstract}
In a recent paper [T. Shang {\it et al.}, \href{http://doi.org/10.1103/PhysRevB.107.224504}{Phys. Rev. B 107, 224504 (2023)]}, the authors study the physical properties of TaReSi compounds having $Ima2$ structure.
This noncentrosymmetric structure is proposed to be the source of topological properties for mentioned compound.
However, for a correct description of topological features, it is important to recognize the correct structure of the compound at low temperature. 
In this Comment, we show that $Ima2$ cannot be realized by TaReSi and is unstable at low temperature.
The $Ima2$ structure contains the soft modes at S (1/2,0,0) point, which leads to the stable structure with $Cm$ symmetry.
Notably, the stable $Cm$ system also has a noncentrosymmetric structure, which can be the actual source of topological properties.
\end{abstract}

\maketitle

\begin{figure}[!b]
\centering
\includegraphics[width=\linewidth]{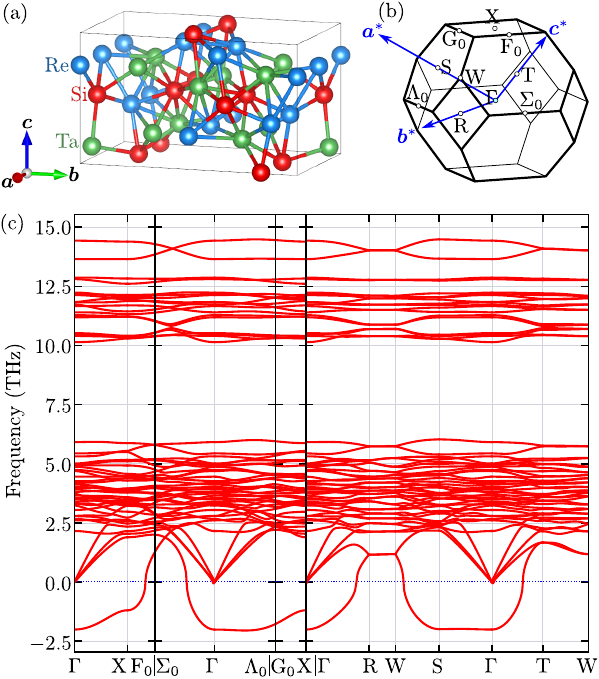}
\caption{
(a) Conventional unit cell of TaReSi with $Ima2$ symmetry  and (b) corresponding the Brillouin zone with their high symmetry points.
(c) Phonon dispersion curves for TaReSi with $Ima2$ symmetry.
\label{fig.phon46}
}
\end{figure}

In the recent paper T.~Shang {\it et al.}~\cite{shang.zhao.23} discusses the properties of TaReSi compound, exhibiting the superconductivity below $T_{c} = 5.5$~K.
The presented theoretical investigation is based on the assumption, that this compound realizes an orthorhombic TiFeSi-like structure, with $Ima2$ symmetry (space group No. 46), presented on Fig.~\ref{fig.phon46}(a).
This statement is supported by powder XRD at normal conditions~\cite{sajilesh.singh.21}, i.e. temperatures much higher than the superconducting state.
Nevertheless, the symmetry of TaReSi can differ in the low temperature regime, which can affect the electronic and topological properties of this compound.
In this Comment, based on the {\it ab initio} calculations, we discuss the stability of TaReSi in the range of low temperatures.

\paragraph*{Computational details.}---
The {\it ab initio} calculations (DFT) are performed using the projector augmented-wave (PAW) potentials~\cite{blochl.94} implemented in the Vienna Ab initio Simulation Package ({\sc Vasp}) code~\cite{kresse.hafner.94,kresse.furthmuller.96,kresse.joubert.99}.
Calculations are made within the generalized gradient approximation (GGA) in the Perdew, Burke, and Ernzerhof (PBE) parameterization~\cite{perdew.burke.96}.
The energy cutoff for the plane-wave expansion was set to $350$~eV.
Optimizations of structural parameters (lattice constants and atomic positions) are performed in the primitive unit cell using the $10 \times 10 \times 6$ {\bf k}--point grid in the Monkhorst--Pack scheme~\cite{monkhorst.pack.76}
As a break of the optimization loop, we take the condition with an energy difference of $10^{-6}$~eV and $10^{-8}$~eV for ionic and electronic degrees of freedom.
The optimized system symmetry was analyzed using {\sc FindSym}~\cite{stoke.hatch.05}.

The dynamical properties were calculated using the direct {\it Parlinski–Li–Kawazoe} method~\cite{parlinski.li.97}, implemented in the {\sc Phonopy} package~\cite{togo.23,togo.chaput.23}. 
Within this method, the interatomic force constants (IFC) are calculated from the forces acting on the atoms after displacement of individual atoms inside a supercell.
We perform these calculations using the supercell containing $2 \times 2 \times 2$ conventional cells, which corresponds to $48$ formula units, and reduced {\bf k}-grid $3 \times 3 \times 3$.

\paragraph*{Stability of TaReSi at low temperatures.}---
The phonon dispersion curves for symmetry $Ima2$ are presented in Fig.~\ref{fig.phon46}(c). 
As we can see, within the phonon spectrum there exist an imaginary soft mode (presented as negative frequencies), which indicated the dynamical instability of TaReSi with $Ima2$ at low temperatures.
However, the atom displacement induced by the soft modes can be used to predict the {\it true} group state of the system.
In this Comment, we present analyses of the possible stable structure of TaReSi at low temperature.

\begin{figure}[!t]
\centering
\includegraphics[width=\linewidth]{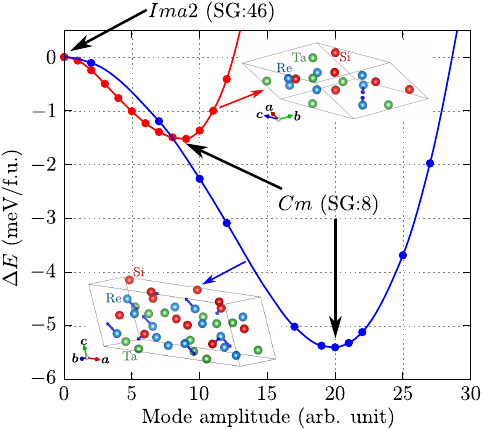}
\caption{
Total energy (per formula unit) of TaReSi as a function of the soft mode amplitude. 
Red and blue color correspond to the displacement generated by the soft mode at $\Gamma$ and S points, respectively (see Fig.~\ref{fig.phon46}).
The relative energies of the initial $Ima2$ and final structures (both with $Cm$ symmetry) are indicated by arrows.
Insets show the crystal structure and atom displacement (blue arrows) introduced by discussed soft modes.
\label{fig.enesoft}
}
\end{figure}

For further analysis, we take soft modes at $\Gamma$ (0,0,0) and S (1/2,0,0) points (i.e. soft modes with the largest magnitude of frequencies).
First, it should be noted, that the soft mode from $\Gamma$ point does not change the size of primitive unit cell, while the one at the S point leads to its doubling along the lattice vector ${\bm a}$.
The displacement of the atoms introduced by the soft modes should lead to new structures with a total energy lower than that of the $Ima2$ structure. 
Indeed, the introduction of atom displacement induced by both soft modes leads to energy lowering, which is clearly seen in Fig.~\ref{fig.enesoft}.

\begin{figure}[!t]
\centering
\includegraphics[width=\linewidth]{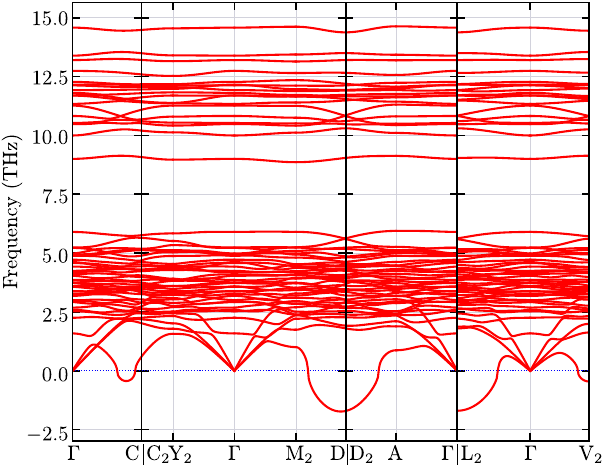}
\caption{
Phonon dispersion curves for TaReSi with $Cm$ symmetry induced by soft mode at $\Gamma$ point (see Fig.~\ref{fig.phon46}).
\label{fig.ph8_gamma}
}
\end{figure}

The symmetries of the structures induced by the both soft modes (before and after structure optimization) are recognized as $Cm$ structures (space group No.~8) [details about the optimized structure can be found in Supplemental Material (SM)~\footnote{See Supplemental Material at [URL will be inserted by publisher] for Crystallographic Information File (CIF) for optimized structures with $Cm$ symmetry.}].
Unfortunately, the displacements of the atoms are related to the entire structure of the compound.
In the case of the soft mode at the $\Gamma$ point (red line in Fig.~\ref{fig.enesoft}), the energy of the system is minimized by the structure when the atoms of Ta, Re, and Si are shifted by $0.025$~\AA, $0.101$~\AA, and $0.046$~\AA, respectively.
Similarly, for the displacement of atoms induced by the soft mode at the S point (blue line in Fig.~\ref{fig.enesoft}), these values are $0.061$~\AA, $0.153$~\AA, and $0.098$~\AA, for Ta, Re, and Si atoms, respectively.

\begin{figure}[!b]
\centering
\includegraphics[width=\linewidth]{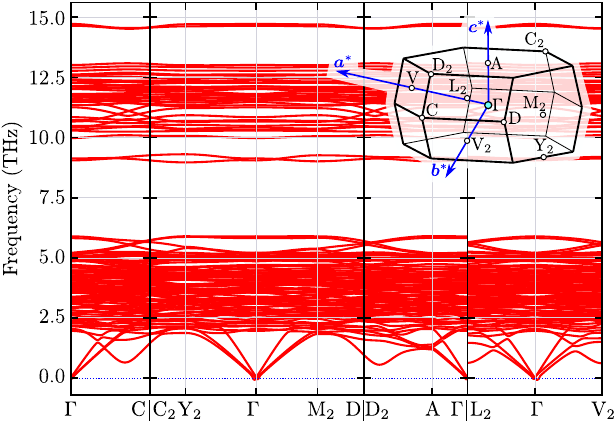}
\caption{
Phonon dispersion curves for TaReSi with $Cm$ symmetry induced by soft mode at S point (see Fig.~\ref{fig.phon46}).
Inset shows the Brillouin zone and its high symmetry points.
\label{fig.ph8_s}
}
\end{figure}

In order to investigate the dynamical stability of ``new'' structure, we reanalyzed the corresponding phonon spectra. 
The phonon dispersion curves for the structure induced by the soft mode at the point $\Gamma$ are presented in Fig.~\ref{fig.ph8_gamma}.
In this case, the phonon spectra still possess the soft mode, which is expected in context of the result presented in Fig.~\ref{fig.enesoft} -- there is structure (induced by the soft mode from the S point) with lower energy.
In the case of this structure, the phonon spectra do not exhibit any imaginary soft modes (Fig.~\ref{fig.ph8_s}), and the structure is stable in the dynamical sense.

In the optimized (stable) structure of TaReSi with $Cm$ symmetry, the atoms are located in 26 non-equivalent positions (see crystal structures in SM~\cite{Note1}): Ta, Re, and Si atoms contain 12, 6, and 8 non-equivalent positions, respectively.
Interestingly, Ta atoms are located only in $2a$ Wyckoff positions, Re atoms only in $4b$ positions, while Si atoms are contained in both $2a$ and $4b$ positions. 
As a result, conventional unit cell contains 24 formula units, which correspond to two primitive unit cells.
Similarly to the case of the previously discussed unstable $Ima2$ structure~\cite{shang.zhao.23}, stable $Cm$ structure is noncentrosymmetric, which allows for the realization of the antisymmetric spin--orbit coupling~\cite{smidman.salamon.17}.

The presented situation can be compared with NbReSi, which is reported as $Ima2$~\cite{sajilesh.motla.22} or $P\bar{6}2m$~\cite{su.shang.21,shang.tay.22,nandi.sasmal.23} structure.
In this case, there is also a soft mode for $Ima2$, and, undoubtedly, the system forms a $P\bar{6}2m$ structure~\cite{basak.ptok.23}.
Contrary to this, TaReSi is recognized as $Ima2$ under normal conditions, whereas the existing soft mode leads to the stable $Cm$ symmetry in case of low temperature regime.


\begin{figure}[!t]
\centering
\includegraphics[width=\linewidth]{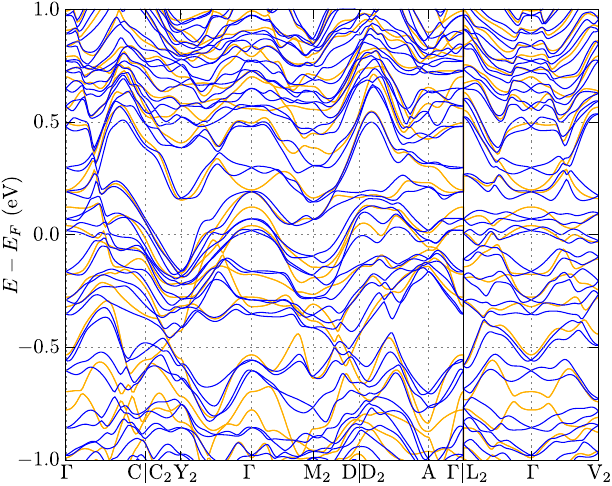}
\caption{
Electronic band structure of TaReSi with $Cm$ symmetry (induced by softmode originated at S point).
Orange and blue lines correspond to the band structure in the spin--orbit coupling absence and presence, respectively. 
\label{fig.el8} 
}
\end{figure}

\begin{figure}[!b]
\centering
\includegraphics[width=\linewidth]{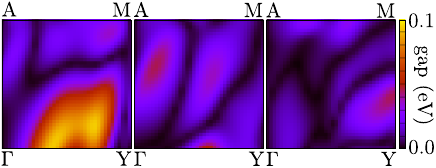}
\caption{
The value of ``gap'' between pairs of bands splitting by the spin--orbit coupling.
Black lines suggest realization of the Kramers nodal lines.
Results for three pairs of bands crossing the Fermi level for plane in the Brillouin zone ($\Gamma$-A-M-Y plane).
\label{fig.kramers}
}
\end{figure}

\paragraph*{Electronic properties.}---
Finally, for the optimized stable $Cm$ structure, we can calculate the electronic band structure (Fig.~\ref{fig.el8}).
The lifting of the band degeneracy by the spin--orbit coupling is well visible (cf. orange and blue lines in Fig.~\ref{fig.el8}).
The calculated band splitting for the $Cm$ structure ($600$~ meV) is much greater than that reported for the $Ima2$ structure ($300$~meV)~\cite{shang.zhao.23}.
In the presence of the spin--orbit coupling (blue lines in Fig.~\ref{fig.el8}), the electronic band structure hosts the double degenerate Weyl points. 
It is worth noting that for the $Ima2$ symmetry~\cite{shang.zhao.23}, there are no Kramers nodal lines along the high-symmetry lines.
Nevertheless, the existence of the mirror symmetry plane $\{ m_{010} | 0 \}$ within $Cm$ symmetry allows for realization of the Kramers nodal lines in the mirror planes. 
Such nodal lines create the close contours between the high symmetry points $\Gamma$, A, M, or Y (represented by the black contours on Fig.~\ref{fig.kramers}).
The vanishing band splitting (no-gap) between pairs of bands splitting by the spin--orbit coupling is clearly visible and is not limited to the high symmetry points~\cite{xie.gao.21}.


\paragraph*{Summary and conclusions.}---
Summarizing, in this Comment, based on the DFT calculations, we establish that the TaReSi in the low temperature range cannot form a stable structure with $Ima2$ symmetry.
This is based on the fact that the phonon spectra calculated for $Ima2$ TaReSi contain the imaginary frequency soft modes. 
The precise examination of the realized symmetry is necessary for discussion of the TaReSi topological properties, which is the main intent of the presented Comment.

Here, we display the calculation suggesting that the $Cm$ structure is more preferable in low temperature regime.
TaReSi with $Cm$ symmetry is still noncentrosymmetric, which is related to the existence of antisymmetric spin--orbit coupling, as claimed in Ref.~\cite{shang.zhao.23}.
Additionally, the spin--orbit coupling strength for $Cm$ symmetry is much larger  than the one reported for $Ima2$, which indeed can support the realization of the topological superconductivity in TaReSi~\cite{shang.zhao.23}.
The absence of inversion symmetry, while preserving mirror symmetry, is a source of Kramers nodal lines~\cite{xie.gao.21}.
In such a case, the vanishing of the band splitting introduced by the spin--orbit coupling forms the closed contours between high symmetry points.
The close vicinity of this band crossing to the Fermi level can be of importance for the topological properties of TaReSi at low temperatures, as reported by experimental observation in Ref.~\cite{shang.zhao.23}.

\paragraph*{Acknowledgments.}---
Some figures in this work were rendered using {\sc Vesta}~\cite{momma.izumi.11} software.
A.P. is grateful to Laboratoire de Physique des Solides in Orsay (CNRS, University Paris Saclay) for hospitality during the work on this project.
This work was supported by National Science Centre (NCN, Poland) under Project No.
2021/43/B/ST3/02166. 


\bibliography{biblio.bib}


\end{document}